\documentstyle[twocolumn,aps]{revtex}

\begin{document}

\draft
\preprint{}
\title{Local Deterministic Transformations of Three-Qubit Pure States}
\author{Federico M. Spedalieri}
\address{Institute for Quantum Information \\ California Institute of Technology, 
MC 452-48, Pasadena, CA 91125, USA}
\date{\today}

\maketitle

\begin{abstract}

The properties of deterministic LOCC transformations of three qubit
pure states are studied. We show that the set of states in the GHZ class
breaks into an infinite number of disjoint classes under this type 
of transformation. 
These classes are characterized by the value of 
a quantity that is invariant under these transformations, 
and is defined in terms of the coefficients of 
a particular canonical form in which only states in the
GHZ class can be expressed. This invariant also imposes a strong constraint
on any POVM that is part of a deterministic protocol. We also 
consider a transformation generated by a local 2-outcome
POVM and study under what conditions it is deterministic, i.e., both
outcomes belong to the same orbit. We prove that for real states it
is always possible to find such a POVM and we discuss analytical 
and numerical evidence that suggests that this result also holds for
complex states. We study the transformation generated in the space of orbits
when one or more parties apply several deterministic POVMs in succession 
and use
these results to give a complete characterization of the real states that can be
obtained from the GHZ state with probability 1. 

\end{abstract}

\newcommand{\pstate}{| \psi \rangle}
\newcommand{\phistate}{| \phi \rangle}
\newcommand{\phizero}{| \phi_0 \rangle}
\newcommand{\phione}{| \phi_1 \rangle}

\section{Introduction}

A very important part of the study of the entangled states of spatially
separated systems, is the study of the transformations that are
possible when using only local operations and classical communication
(LOCC), since it allows us to classify entangled states and it can be
used as one way of quantifying this resource. Two states that 
are related by local unitary transformations are considered equivalent
as far as entanglement is concerned, since both states can be obtained from each
other and local operations cannot increase entanglement. The action of 
the group of local unitaries breaks
the space of states into orbits \cite{multpent}. 
Then, to transform a pure state into 
another state in a different orbit by local
operations, we need to allow each party to apply a local generalized measurement,
i.e., a POVM , on her part of the state.

For bipartite pure states, the problem of deterministically transforming 
a state into another has been solved
by Nielsen \cite{nielsen}, who gave necessary and sufficient conditions for 
a given transformation to be achievable with probability 1. Later Vidal
\cite{probtran} extended this result by calculating the maximal 
probability of success of
any LOCC transformation of bipartite pure states.
For more than two parties, this problem is still unsolved. 
The bipartite case seems to be very special due to the existence
of the Schmidt decomposition. Any pure bipartite state can  be transformed
by applying local unitaries into a state of the form
\begin{equation}
\pstate = \sum_i^n \lambda_i | i i \rangle ,
\end{equation}
where the $\lambda_i$ are positive real numbers, $| i i \rangle
=|i\rangle_A \otimes |i\rangle_B$ and $\{|i\rangle\}$
are orthonormal vectors on each subsystem.  This greatly simplifies 
the analysis
of LOCC transformations: it gives a canonical expression
for states in a given orbit, and 
allows the reduction of an arbitrary LOCC protocol to a protocol in which
one party applies local unitaries and local POVMs, and the other party only
has to apply a local unitary, conditional on the results obtained
by the first party \cite{redprot}. For multipartite states with three 
parties or more
there is no known  reduction of LOCC protocols.

For a system of three qubits, several Schmidt-like decompositions
have been proposed \cite{genschmidt,tacjal}, all based on the idea 
of using local unitaries
to get rid of as many coefficients as possible. One interesting property
that emerges from these decompositions is that in general it is not possible
to make all the coefficients real. In particular there are states that
have at least one coefficient that is complex for any local basis, and this 
has as a consequence that these
states are not locally unitarily equivalent to their conjugates (the states
obtained by taking the complex conjugate of the coefficients).  This 
contrasts with the bipartite
case in which, since the Schmidt decomposition has only real coefficients, 
every state is in the same orbit as its conjugate. 

A POVM applied to a state has, in general, outcomes that belong to
different orbits. However, a protocol that transforms a state into
another with probability 1, \textit{has to include at least one} POVM for
which all outcomes are in the same orbit. For instance, this has
to be the case for the last POVM of the
protocol: if its outcomes are not in the same orbit,
then the protocol has not achieved the transformation with probability 1.
We will call a POVM with 
this property a \textit{deterministic} POVM, because we can use such a POVM 
and suitable local unitaries,
to obtain any state in the orbit of the outcomes with probability 1,
attaining a deterministic transformation. Since any local POVM can 
be replaced by a sequence of 2-outcome POVMs, it is then interesting
to study the case of a deterministic 2-outcome POVM.

In this paper we will study some properties of deterministic
LOCC protocols and deterministic POVMs applied to 3-qubit pure states.
We will only be interested in transformations between states
that have genuine tripartite entanglement (i.e., all three
reduced density matrices have rank 2), since 
other cases can be reduced to the bipartite case.
In Section II, we prove that a certain function of the states is
invariant under any deterministic LOCC protocol and show
that this imposes a constraint on the local POVMs that can be a part 
of a deterministic transformation. We also show that the set of states in the GHZ class 
breaks into an infinite number of
disjoint subclasses under this particular type of transformation. In Section III, 
we study the particular
case of a 2-outcome deterministic POVM, and discuss what are the conditions
for its existence. We prove that such a POVM can always be found
for real states, and present some evidence that the same situation holds
for complex states. In Section IV we analyze the transformation
in the space of orbits. In Section V we study the case of
the GHZ state and give a complete characterization of all the states with real
coefficients that
can be obtained deterministically from it. Finally, the conclusions
are presented in Section VI.

\section{General properties of LOCC transformations of 3-qubit states}

Pure states of three qubits with 3-particle entanglement are divided
in two inequivalent classes: the GHZ class and the W class \cite{wghz}. 
They have the property that any local POVM applied to a state in a
given class, can only have as outcomes states in the same class. In 
particular,
states in the W class can always be transformed by local
unitary operations, into a state with real coefficients. In this
paper we will call a state ``real'' if it is
locally unitarily equivalent (LUeq) to a state with
real coefficients. States in the GHZ class
can be either real or complex.

Any state in the GHZ class is LUeq to a state of the form \cite{wghz}
\begin{equation}
\label{ghzdec}
\pstate = \mu |0 0 0 \rangle + \nu e^{i \gamma} |\varphi_A \rangle 
 |\varphi_B \rangle  |\varphi_C \rangle,
\end{equation}
where $\mu \geq \nu > 0$ are real numbers, $\gamma \in [0,2 \pi )$ and 
$ |\varphi_X \rangle  = \cos \delta_X |0\rangle
+\sin \delta_X |1\rangle$ with $\delta_X \in (0,\frac{\pi}{2}]$ and
$X = A,B,C$. We will assume that the state $\pstate$ is normalized, so
only five of the six parameters in (\ref{ghzdec}) are independent. If we 
write $\pstate = |\mu \rangle +|\nu \rangle$
where $|\mu \rangle$ and $|\nu \rangle$ correspond to the first and second
term in (\ref{ghzdec}) respectively, we can construct the invariant
\begin{equation}
\label{munu}
\Omega (\pstate)=\langle \mu | \nu \rangle = \mu \nu e^{i \gamma} \cos \delta_A 
\cos \delta_B \cos \delta_C.
\end{equation}
If $\mu = \nu$, the sign of the phase $\gamma$ is not well defined, since in
this case there is an ambiguity with respect to which product state in 
(\ref{ghzdec}) is $|\mu\rangle$, and hence we can interchange $|\mu\rangle$ and
$|\nu\rangle$ by local unitaries, and transform the state into its conjugate,
which changes the sign of $\gamma$. 
As shown in \cite{acin1} this means that the
state is real, although we need to use complex coefficients if we want
to write it in the particular form given by (\ref{ghzdec}). 
Aside from this ambiguity, this decomposition is unique.
If $\mu > \nu$ then the state $\pstate$ is complex if and only if ${\mathrm Im}(\Omega 
(\pstate)) \neq 0$. If  ${\mathrm Im}(\Omega (\pstate))=0$, then either $\gamma$ 
is equal to $0$ or $\pi$ (and in both cases all the coefficients are real, 
so the state is real), or
$\delta_X =\frac{\pi}{2}$ for some $X$. If this is the case, we can get rid 
of the phase by applying the local unitary
\begin{equation}
U = \left( \begin{array}{cc} 1 & 0 \\ 0 & e^{-i \gamma}
                   \end{array} \right),
\end{equation}
to party $X$, which makes all the coefficients real.

Let $\{A_i \}, i=1,\ldots,n$ represent a local POVM applied by Alice.
If we apply it to a state $\pstate$ we
can write the normalized outcomes as $| \phi_i \rangle = 
q_i^{-\frac{1}{2}} A_i \otimes \mathbf{1} \otimes \mathbf{1} \pstate$, where
$q_i = \langle \psi | A_i^\dag A_i \otimes \mathbf{1} \otimes \mathbf{1} 
\pstate$ is the probability of outcome $i$. Let's consider the case in which
none of the operators $A_i$ corresponds to a projective measurement (i.e.,
they all have rank 2). If we apply this POVM to a state with
genuine tripartite entanglement, all the outcomes will still
have 3-particle entanglement. To understand why this is true, suppose that there
is an operator $A_j$ of the POVM such that its corresponding outcome $|\phi_j
\rangle$ has no
3-particle entanglement. Then $|\phi_j \rangle$ has to be the product of
a pure state of one of the parties, let's call it $X$, and a pure state 
(possibly entangled) of the remaining two parties, so party $X$ is
completely disentangled from the other two. Since we are assuming 
that $A_j$ is invertible (it is a rank two, 2 by 2 matrix), we can
construct a local POVM with operators $\{A_j^{-1}, 
\sqrt{{\mathbf 1}-(A_j^{-1})^\dag A_j^{-1}}\}$
that when applied to $|\phi_j \rangle$ has at least one outcome that has
3-particle entanglement (the one corresponding to $A_j^{-1} \otimes {\mathbf 1}
\otimes {\mathbf 1} |\phi_j \rangle$), that
occurs with nonzero probability (because $(A_j^{-1})^\dag A_j^{-1}$ also has
rank two). Then we would have a protocol that with finite probability and only 
applying local operations, allows us to create entanglement between party $X$ and
the other two, starting from a state in which party $X$ was disentangled, and
this is clearly not possible.

Let's consider a state $\pstate$ in the
GHZ class and let Alice apply a local POVM to it. Then all the 
outcomes $| \phi_i \rangle$ have to be
in the GHZ class too, so we know that we can apply local unitaries to 
them such that
\begin{equation}
(U_{A (i)} \otimes U_{B (i)} \otimes U_{C (i)}) | \phi_i \rangle = 
| \mu_i \rangle + | \nu_i \rangle,  
\end{equation}
where
\begin{eqnarray}
\label{muinui}
| \mu_i \rangle & = & \mu_i |0 0 0 \rangle \nonumber \\
| \nu_i \rangle & = & \nu_i e^{i \gamma_i} |\varphi_{A (i)} \rangle 
 |\varphi_{B (i)} \rangle  |\varphi_{C (i)} \rangle.
\end{eqnarray}
Since $| \mu_i \rangle , | \nu_i \rangle$ and $| \mu \rangle , | \nu \rangle$
are product states, and the action of the POVM and any local unitaries
is still local, for every outcome $i$ we must have either
\begin{eqnarray}
\label{prodcorr1}
\sqrt{q_i}|\mu_i \rangle & = & (U_{A (i)} \otimes U_{B (i)} \otimes U_{C (i)})
( A_i \otimes \mathbf{1} \otimes \mathbf{1}) | \mu \rangle
\nonumber \\
\sqrt{q_i}|\nu_i \rangle & = & (U_{A (i)} \otimes U_{B (i)} \otimes U_{C (i)})
( A_i \otimes \mathbf{1} \otimes \mathbf{1}) | \nu \rangle, 
\end{eqnarray}
or
\begin{eqnarray}
\label{prodcorr2}
\sqrt{q_i}|\mu_i \rangle & = & (U_{A (i)} \otimes U_{B (i)} \otimes U_{C (i)})
( A_i \otimes \mathbf{1} \otimes \mathbf{1}) | \nu \rangle
\nonumber \\
\sqrt{q_i}|\nu_i \rangle & = & (U_{A (i)} \otimes U_{B (i)} \otimes U_{C (i)})
( A_i \otimes \mathbf{1} \otimes \mathbf{1}) | \mu \rangle. 
\end{eqnarray}
To decide which one is the case, we note that decomposition (\ref{ghzdec})
requires that $\mu_i \geq \nu_i$, and $\mu_i , \nu_i$ are the norms of the
states $|\mu_i\rangle$ and $|\nu_i\rangle$ respectively. Then, if 
$\langle \mu | A_i^\dag A_i \otimes {\mathbf{1}} \otimes {\mathbf{1}}
 |\mu\rangle \geq
\langle \nu | A_i^\dag A_i \otimes {\mathbf{1}} \otimes {\mathbf{1}} |\nu\rangle$ 
(which is equivalent to $\langle \mu_i |\mu_i\rangle \geq
\langle \nu_i |\nu_i\rangle$), we have that (\ref{prodcorr1}) must hold. Otherwise,
(\ref{prodcorr2}) holds. Using $\sum_i A_i^\dag A_i = \mathbf{1}$, we can then write
\begin{eqnarray}
\label{phaseave}
{\mathrm{Re}}(\Omega(\pstate)) = \langle \mu | \nu \rangle + 
\langle \nu | \mu \rangle & = & \sum_i q_i (\langle \mu_i 
| \nu_i \rangle + \langle \nu_i | \mu_i \rangle) \nonumber \\
 & = & \sum_i q_i {\mathrm{Re}}(\Omega(|\phi_i\rangle)). 
\end{eqnarray}
This result is due to Vidal \cite{vidpriv}. It puts a strong constraint on
deterministic LOCC protocols, as we show in the following theorem. 

\newtheorem{thm1}{Theorem} 
\begin{thm1}
Let $\pstate$ and $|\xi\rangle$ be two states in the GHZ class and assume 
there is a LOCC
protocol that transforms $\pstate$ into $|\xi\rangle$ with probability 1.
Then, 
\begin{equation}
{\mathrm{Re}}(\Omega(\pstate)) = {\mathrm{Re}}(\Omega(|\xi\rangle)),
\end{equation} 
i.e., the quantity ${\mathrm{Re}}(\Omega)$ is invariant
under deterministic LOCC transformations.
Furthermore, it must be invariant
for every local POVM in the protocol, that is, if the POVM is applied to a
state $|\chi\rangle$ and has outcomes $|\phi_i\rangle$, then
\begin{equation}
{\mathrm{Re}}(\Omega(|\chi\rangle)) = {\mathrm{Re}}(\Omega(|\phi_i\rangle)),
\end{equation}
for all $i$.
\end{thm1}
\textbf{Proof:} The most general LOCC protocol is a sequence of local
unitaries, local POVMs and classical communication between all the
parties. Local unitaries cannot change ${\mathrm{Re}}(\Omega)$
because $\Omega(\pstate)$ is an invariant of the orbit. Thus, it can only
be changed by applying POVMs. Consider the first POVM
of the protocol, that takes the state $\pstate$ into one of its possible 
outcomes $|\phi_i\rangle$, each occurring with 
probability $q_i$. Then, according to equation (\ref{phaseave})
(and because $q_i>0$), either
all outcomes $|\phi_i\rangle$ satisfy
${\mathrm{Re}}(\Omega(|\phi_i\rangle)) = {\mathrm{Re}}(\Omega(\pstate))$ or there 
are at least two outcomes
$|\phi_1\rangle$ and $|\phi_2\rangle$ that satisfy ${\mathrm{Re}}(\Omega(|\phi_1
\rangle))<{\mathrm{Re}}(\Omega(\pstate))<{\mathrm{Re}}(\Omega(|\phi_2\rangle))$. 
It is easy to see that in the latter case, at any stage
in the protocol, we will have two outcomes $|\phi_j\rangle$ and $|\phi_k\rangle$
that will satisfy ${\mathrm{Re}}(\Omega(|\phi_j\rangle))<{\mathrm{Re}}(\Omega(|\phi_k
\rangle))$. 
This will be true in particular
for the last stage of the protocol. But that would mean that $|\phi_j\rangle$ and 
 $|\phi_k\rangle$ are in different orbits (because $\Omega$ is invariant
under local unitaries), and that contradicts the fact that the protocol is
deterministic. Thus, the only possibility is that all the outcomes
of the first POVM have the same value of ${\mathrm{Re}}(\Omega)$. We can apply
exactly the same reasoning to all the POVMs in the protocol, and then
conclude that all the final outcomes satisfy ${\mathrm{Re}}(\Omega (|\phi_i\rangle))
 = {\mathrm{Re}}(\Omega(\pstate))$.
Since this is a deterministic protocol that transforms $\pstate$ into
$|\xi\rangle$, then all these outcomes should be in the same orbit as
$|\xi\rangle$, and so we have ${\mathrm{Re}}(\Omega (|\xi\rangle))={\mathrm{Re}}(\Omega 
(|\phi_i\rangle))={\mathrm{Re}}(\Omega(\pstate))$.$\Box$

This theorem tells us that under deterministic LOCC transformations
the class of GHZ states breaks into an infinite number of subclasses
that are labeled by the real part of the complex invariant $\Omega$. Two states in
different subclasses cannot be transformed one into the other with probability
1 by means of local operations and classical communication. From equation 
(\ref{munu})
and from the range of the parameters, we see that the set of these subclasses
is isomorphic to the open segment $(-\frac{1}{2},\frac{1}{2})$. The subclass that 
contains the GHZ state $|GHZ\rangle = 
\frac{1}{\sqrt{2}} (|000\rangle +|111\rangle)$, corresponds to the center
of the segment, and it is defined by ${\mathrm{Re}}(\Omega)=0$. Note that all
subclasses contain both real and complex states.

This result gives a broad description of how a state
can be transformed in the space of orbits with probability 1.
Tighter constraints can be obtained from studying the
behavior of the entanglement monotones \cite{em}, which usually
introduce some necessary conditions that must be satisfied in order
for a transformation to be possible to be implemented locally.
To find sufficient conditions we have to be able to show
that a protocol exists that accomplishes the transformation. 
A first step in that direction is to study deterministic
POVMs.

\section{Deterministic 2-outcome POVM}

In this section we will study under what conditions a 2-outcome
POVM is a deterministic POVM (i.e., both outcomes are in the
same orbit). A general 3-qubit state can be written
\begin{equation}
\label{state}
  \pstate = \sum_{i j k =0}^1 t_{i j k} | i j k \rangle.
\end{equation}
Following \cite{tacjal}, we can define matrices $T_0$ and $T_1$, where
\begin{equation}
\label{Tmatrix}
  (T_i)_{ j k} = t_{i j k}.
\end{equation}
The group of Local Unitary (LU) transformations of three qubits is locally 
isomorphic (i.e., has the same Lie algebra) to $U(1)\times [ SU(2)]^3$. 
Under a LU transformation performed only by Bob and Charlie with matrices 
$U_B$ and $U_C$, the matrices $T$ transform according to
\begin{equation}
\label{transfTBC}
  T_i \rightarrow U_B T_i U_C,
\end{equation}
while if the transformation is performed by Alice, we have
\begin{eqnarray}
\label{transfTA}
  T_0 & \rightarrow & u_{0 0}^A T_0 + u_{0 1}^A T_1 \nonumber \\
  T_1 & \rightarrow & u_{1 0}^A T_0 + u_{1 1}^A T_1, 
\end{eqnarray}
where $u_{i j}^A$ are the matrix elements of $U_A$. 

We know \cite{3qinvariants,bob} that the orbits of 3-qubit states can be 
parametrized with 5 continuous invariants plus a discrete invariant, 
since in general a 3-qubit state is not LUeq to its complex conjugate. 
There are many ways of choosing these invariants \cite{wghz,genschmidt,tacjal}. 
In this paper we will use the following set

\begin{eqnarray}
\label{invariants}
I_1 & = & \sum_{i j k m p q} t_{k i j} t_{m i j}^\ast t_{m p q} t_{k p q}^\ast = tr \rho_A^2  \nonumber \\
I_2 & = & \sum_{i j k m p q} t_{i k j} t_{i m j}^\ast t_{p m q} t_{p k q}^\ast = tr \rho_B^2 \nonumber \\
I_3 & = & \sum_{i j k m p q} t_{i j k} t_{i j m}^\ast t_{p q m} t_{p q k}^\ast = tr \rho_C^2 \nonumber \\
I_4 & = & |\sum_{i j k l m n o p q r s t} t_{i j k}  t_{l m n}  t_{o p q}  t_{r s t} \epsilon_{i l} \epsilon_{o r} \epsilon_{j m} \epsilon_{p s} \epsilon_{k q} \epsilon_{n t} | \nonumber \\
I_5 & = &  \sum_{i j k l m n o p q} t_{i j k} t_{i l m}^\ast t_{n l o} t_{p j o}^\ast  t_{p q m} t_{n q k}^\ast ,
\end{eqnarray}
where $\epsilon_{i j}$ is the antisymmetric symbol and all the indices
 are summed from 0 to 1. $I_4$ is the 3-tangle introduced in \cite{tangle}. 
As shown in
\cite{3qinvariants} these 5 invariants are algebraically independent. However, 
since they are all real and invariant under complex conjugation of the
coefficients $t_{ijk}$, they cannot distinguish between a 
state and its conjugate. To fix this ambiguity we use the 
complex invariant \cite{I6bob}

\begin{eqnarray}
\label{complexinv}
  I_6 & = & \sum_{i_l j_l k_l} t_{i_1 j_1 k_1}  t_{i_2 j_2 k_2} t_{i_3 j_3 k_3} t_{i_4 j_4 k_4} t_{i_5 j_5 k_5} t_{i_6 j_6 k_6} \times \nonumber \\
  & & \ \ \ \ \ t_{i_1 j_1 k_3}^\ast t_{i_2 j_2 k_4}^\ast t_{i_3 j_4 k_5}^\ast t_{i_4 j_3 k_1}^\ast t_{i_5 j_6 k_2}^\ast t_{i_6 j_5 k_6}^\ast,
\end{eqnarray}
 where again all indices are summed from 0 to 1. To completely specify an orbit
we need the value of $I_1$ through $I_5$ plus the sign of the imaginary
part of $I_6$. It is worth noting that $1-I_1$ , $1-I_2$ , $1-I_3$ and $I_4$ 
are decreasing entanglement monotones, while $I_5$ is not an entanglement
monotone \cite{bobI5}.

We will consider the case of a 2-outcome POVM applied by Alice on
a pure state $\pstate$ of three qubits. The most general  
POVM is given by the operators $A_0$ and $A_1$, where
\begin{eqnarray}
\label{2outpovm}
  A_0 & = & V_0 \left( \begin{array}{cc} \sqrt{x} & 0 \\ 0 & \sqrt{y}
                   \end{array} \right) U  \nonumber \\
  A_1 & = & V_1  \left( \begin{array}{cc} \sqrt{1-x} & 0 \\ 0 & \sqrt{1-y}
                   \end{array} \right) U ,
\end{eqnarray}
where $V_0$, $V_1$ and $U$ are unitary matrices \cite{svd}, and \hbox{$0 \leq x , 
y \leq 1$.}
It is easy to see that they satisfy \hbox{$A_0^\dag A_0 + A_1^\dag A_1 = 
\mathbf{1}$,} where $\mathbf{1}$ is the identity matrix. When we apply
this POVM to a state $\pstate$, we obtain two outcomes $\phizero$ and 
$\phione$ given by
\begin{equation}
\label{povmoutcome}
  | \phi_i \rangle = \frac{1}{\sqrt{q_i}}(A_i \otimes {\mathbf 1} \otimes \
{\mathbf 1} 
\pstate) \ \ \ \ i=0,1,
\end{equation} 
where $q_i$ is the probability
of outcome $i$. From (\ref{2outpovm}) and (\ref{povmoutcome}) we can see that
the action of this POVM on $\pstate$ is equivalent to applying a unitary
tranformation first given by $U$, applying a diagonal and real POVM and
finally applying a unitary $V_i$ conditional on the outcome of the POVM. 
This last local unitary cannot change the orbit of the outcome $| \phi_i
\rangle$. Since we are considering two states in the same orbit to be
equivalent, we can take this unitary to be the identity without loss of 
generality.

Let us consider first the case in which $U=\mathbf{1}$. Then
both elements of the POVM reduce to real and diagonal matrices
\begin{equation}
\label{diagpovm}
  E_0 = \left( \begin{array}{cc} \sqrt{x} & 0 \\ 0 & \sqrt{y} \end{array}
        \right) \ \  , \ \ 
  E_1 = \left( \begin{array}{cc} \sqrt{1-x} & 0 \\ 0 & \sqrt{1-y} \end{array}
        \right). 
\end{equation}
From now on, we will take $0 < x ,y < 1$, since when $x$ or $y$ are 
equal to zero or one, the POVM becomes a projective measurement, which destroys
three particle entanglement.  We can write explicit expressions for both 
outcomes of the POVM

\begin{eqnarray}
\label{outcomes}
  \phizero & = & \frac{1}{\sqrt{q_0}} \sum_{j k} (\sqrt{x} \ t_{0 j k} 
|0 j k \rangle + \sqrt{y} \ t_{1 j k} |1 j k \rangle ) \nonumber \\
  \phione & = & \frac{1}{\sqrt{q_1}} \sum_{j k} (\sqrt{1-x}\ t_{0 j k} 
|0 j k \rangle + \sqrt{1-y} \ t_{1 j k} |1 j k \rangle ).
\end{eqnarray}

Now we calculate the invariants $I_1$ through $I_5$ for $\phizero$ as a 
function of $x$ and $y$

\begin{eqnarray}
\label{invxy}
I_1 (x,y) & = & \frac{x^2 a^2 + 2 x y ~ Tr[T_0 T_1^\dag ]~
Tr[T_1 T_0^\dag ] + y^2 b^2}{ ( a x + b y )^2} 
\nonumber \\
I_2 (x,y) & = & \frac{x^2 F_0  + 2 x y ~ Tr[T_0 T_0^\dag T_1 
T_1^\dag ] + y^2 F_1}{ ( a x + b y )^2} 
\nonumber \\
I_3 (x,y) & = & \frac{x^2 F_0  + 2 x y ~ Tr[T_0 T_1^\dag T_1 
T_0^\dag ] + y^2 F_1}{ ( a x + b y )^2} 
\nonumber \\
I_4 (x,y) & = & \frac{x  y ~ I_4(\pstate)}{ ( a x + b y )^2} \nonumber \\
I_5 (x,y) & = & \frac{x^3 ~ G_{0 0}+ 3 x^2 y ~ G_{0 1} + 3 x y^2  ~G_{1 0} + 
y^3 ~ G_{1 1}}
{ ( a x + b y )^3}, 
\end{eqnarray}
where the matrices $T_i$ are as defined in (\ref{Tmatrix}), \hbox{$a
=Tr[T_0 T_0^\dag]$}, $b=Tr[T_1 T_1^\dag ]$, $a + b = 1$ for a normalized 
$\pstate$, 
$F_i = Tr[(T_i T_i^\dag)^2]$ and
$G_{i j}= Tr[T_i T_j^\dag T_i T_i^\dag T_j T_i^\dag]$. The
invariants for $\phione$ are obtained from (\ref{invxy}) by replacing $x$ by
$1-x$ and $y$ by $1-y$. For the two outcomes to be in the same orbit,
we need the five invariants  to take the same values for both states, i.e.,
\begin{equation}
\label{equalinv}
  I_i (x,y) = I_i (1-x,1-y)  \ \ \ \  i=1,\ldots ,5.
\end{equation}
If these conditions are satisfied, then either $\phizero$ is LUeq to 
$\phione$, or
$\phizero$ is LUeq to $\phione^\ast$. To determine which one is the case, 
we need to
calculate the sign of the imaginary part of the complex invariant
$I_6$. For now, let us concentrate on the equations in (\ref{equalinv}).
These equations have a common solution with $0 < x , y < 1$ if and
only if the following conditions are satisfied (see appendix)
\begin{eqnarray}
  a^2 \ Tr[ (T_1 T_1^\dag )^2] &  = & b^2 \  
Tr[ (T_0 T_0^\dag )^2]   \label{cond1} \\
  a \ Tr[ T_1 T_0^\dag T_1 T_1^\dag T_0 T_1^\dag ] & = &
   b \ Tr[ T_0 T_1^\dag T_0 T_0^\dag T_1 T_0^\dag ]  
\label{cond2} \\
 a^2 x (1-x) & = & b^2 y (1-y) \label{condpovm}. 
\end{eqnarray}
Furthermore, the solution satisfies $I_5 (| \phi_i \rangle ) < I_5 (\pstate)$.
This
is worth noting because $I_5$ is not an entanglement monotone, but
behaves monotonically under this particular class of POVMs. Equations
(\ref{cond1}) and (\ref{cond2}) are real valued polynomial constraints on the
coefficients of the state, and in general are not satisfied for an
arbitrary state. From 
(\ref{transfTBC}) and (\ref{transfTA}) 
we can see that these constraints are invariant under LU transformations 
applied by Bob and Charlie, while they are not invariant under 
local unitaries by Alice. Equation (\ref{condpovm}) is a constraint
on the parameters of the POVM that depends on the state we are
transforming. 

Now let $U$ be any unitary matrix, so our POVM takes the
form $\{E_0 U , E_1 U\}$, with
$E_0 , E_1$ given by (\ref{diagpovm}). This is equivalent to applying
the local unitary $U$ to Alice's part of the state, followed by a 
diagonal POVM, and we know the conditions that need to be satisfied
in this last stage. So we can reduce the problem to finding a local
unitary performed by Alice that would transform the original state
$\pstate$ into a state that satisfies (\ref{cond1}) and (\ref{cond2}).
Then we can choose a POVM that satisfies (\ref{condpovm}), where now
$a$ and $b$ are calculated using the coefficients of the transformed
state $U \otimes \mathbf{1} \otimes \mathbf{1} \pstate$. We will consider
the cases of real and complex states separately.

\subsection{Real States}

To characterize the orbit of a real state $\pstate$ we only need four 
parameters instead of the five needed for an arbitrary state. First, note
that, by our definition, any real state can be transformed by means of local 
unitary transformations, into
a state with only real coefficients. Of the (at most) eight coefficients of
this state,
only seven are  independent if we are considering a 
normalized state, and we can get rid of three more by applying local
real unitary (orthogonal) transformations on each of the 3 qubits.
Since $I_i , i=1,\ldots,4$ are algebraically independent, we can use
this set to parametrize the orbits of real states. This greatly 
simplifies our analysis because, as seen in the appendix, (\ref{cond1}) 
is enough to assure
that $I_i , i=1,\ldots,4$ have the same values for both outcomes
of our POVM. So, given a real state, we need to find a $U$ such that
$| \psi' \rangle = U \otimes \mathbf{1} \otimes \mathbf{1} \pstate$ 
satisfies (\ref{cond1}).
Let 
\begin{equation}
\label{realU}
  U (\alpha) =  \left( \begin{array}{cc} \cos \alpha & \sin \alpha
 \\ - \sin 
\alpha & \cos \alpha  \end{array} \right).
\end{equation}
In terms of the matrices $T_i$, this transformation can be written
\begin{eqnarray}
\label{realtransfT}
  T_0' & = &  \cos \alpha \ T_0 + \sin \alpha \ T_1 
\nonumber \\
  T_1' & = & - \sin \alpha \ T_0 + \cos \alpha \ T_1 .
\end{eqnarray}
If we plug this into (\ref{cond1}), take out a common factor $\cos ^8 \alpha$,
introduce the variable $z = \tan \alpha$ and move all terms to one
side, we can write (\ref{cond1})
as polynomial $p_1 (z)$ of degree 8 with real coefficients, of the form
\begin{eqnarray}
\label{realp1}
   p_1 (z) & = & A (1 - z^8) + B (z + z^7) + C (z^2 - z^6) + \nonumber \\
           &   & \ \ \ \ \ \ \ \ \ \ \ \ \ \   + D (z^3 +z^5) = 0 .
\end{eqnarray}
If $z_0$ is a real root of $p_1$, then $U(\alpha_0)$, with $\alpha_0 = 
\arctan (z_0)$ is the unitary matrix we are looking for. Now it's easy
to check that $p_1 (1) = -p_1 (-1)$, so $p_1$ has at least one real root
in $[-1,1]$, which means that we can always find a unitary $U$, such 
that $| \psi' \rangle = U \otimes \mathbf{1} \otimes \mathbf{1} \pstate$
satisfies (\ref{cond1}). Now we can apply to $| \psi' \rangle$ a diagonal
POVM that satisfies (\ref{condpovm}), and we are certain that both outcomes
have the same values of the four invariants $I_i , i=1,\ldots,4$. But in the
case of real states this is enough to completely specify the orbit, because
since $\pstate$ is real, so is $ | \psi' \rangle$ because $U$ was
chosen to be real, and the outcomes of the POVM, $\phizero$ and $\phione$, are
also real, because the POVM itself is real. In this case we don't
have to worry about the value of the complex invariant. Finally,
since $\phizero$ and $\phione$ are in the same orbit, we can apply local
unitaries to transform them into any state in the same orbit. So the
net result of this protocol is to transform any state in the orbit 
of $\pstate$ into any state in the orbit of $\phizero$, with probability 1.
The results presented so far show that for any real state, there is 
some set of orbits that can be reached deterministically from that state,
although we haven't yet characterized this set. We will discuss this
problem in Section IV. 

\subsection{Complex states}

The analysis of the complex states turns out to be more
complicated, because now we need to find $U$ such that $ | \psi' \rangle$
satisfies \textit{both} (\ref{cond1}) and (\ref{cond2}). We can 
write any unitary $U$ as 
\begin{equation}
\label{unitdec}
 e^{i \phi} \left( \begin{array}{cc} e^{i \beta} & 0 \\ 0 & e^{- i \beta} 
\end{array}
        \right)  \left( \begin{array}{cc} \cos \alpha & \sin \alpha \\ - \sin 
\alpha & \cos \alpha \end{array}
        \right)  \left( \begin{array}{cc} e^{i \zeta} & 0 \\ 0 & e^{- i \zeta}
 \end{array}
        \right).
\end{equation}
The phase and the matrix on the left will commute with the diagonal 
matrices of the POVM, so their action is equivalent to applying
a local unitary to the outcomes of the POVM. But we know that this
action will not change the orbit of the outcome state, so we
can fix them to be the identity. So $U$ will take the state $\pstate$ 
with matrices
$T_0$ and $T_1$ to a state $| \psi ' \rangle$ with matrices
\begin{eqnarray}
\label{transfT}  
  T_0' & = & \cos \alpha \  e^{i \zeta} \ T_0 + \sin \alpha \ e^{- i \zeta} \ 
T_1 
\nonumber \\
  T_1' & = & - \sin \alpha \ e^{i \zeta} \ T_0 + \cos \alpha \ e^{- i \zeta} \
T_1.
\end{eqnarray}
We can substitute (\ref{transfT}) into the homogeneous form of (\ref{cond1}) and 
(\ref{cond2}), again divide by
$\cos^8 \alpha$ and introduce the variable $z = \tan \alpha$, so both 
conditions
are expressed as polynomials in $z$ equal to zero, with real coefficients, 
of the form
\begin{eqnarray}
\label{polycond}
  p_i (z) & =& A_i (1 - z^8) + B_i (z + z^7) + C_i (z^2 - z^6) + \nonumber \\
          &  & \ \ \ \ \ \ \ \ \ \ \ + D_i (z^3 +z^5) = 0 \ \ \ \ i = 1,2\ ,
\end{eqnarray}
with the coefficients given by
\begin{eqnarray}
\label{coeff}
A_i & = & a_{0 i} \nonumber \\
B_i & = & b_{1 i} \cos (2 \zeta) + b_{2 i} \sin (2 \zeta) \nonumber \\
C_i & = & c_{0 i} + c_{1 i} \cos (4 \zeta) + c_{2 i} \sin (4 \zeta) 
\nonumber \\
D_i & = & d_{1 i} \cos (2 \zeta) + d_{2 i} \sin (2 \zeta)+ d_{3 i} 
\cos (6 \zeta) + \nonumber \\ & & \ \ \ \ \ \ \ \ \ \ \ \ 
+ d_{4 i} \sin (6 \zeta),  
\end{eqnarray}
where $a_{0 i} , b_{j i} , c_{j i} , d_{j i}$ are real valued polynomials 
on the
coefficients of $\pstate$, whose exact form can be
computed from regrouping the terms obtained after substituting (\ref{transfT})
into (\ref{cond1}) and (\ref{cond2}). 

Finding a local unitary performed by Alice on $\pstate$ that would yield
a state that satisfies (\ref{cond1}) and (\ref{cond2}), is equivalent to 
finding values  $z$ and $\zeta$ (which parametrize the unitary) such
that both polynomials $p_1$ and $p_2$ vanish. We can think of $\zeta$ as
a parameter for these polynomials, and what we are looking for is 
a value of $\zeta$ such that $p_1$ and $p_2$ have a common real root.

The polynomials $p_i$ have certain useful symmetries. First of all, because
their coefficients are real, complex roots appear in conjugate pairs. Also,
because of the particular symmetry of the coefficients (i.e., the 
coefficient of $z^8$ is equal to minus the independent term, the
coefficient of $z^7$ is equal to the coefficient of $z$, and so on), if
$z_0$ is a root of $p_i$, so is $- \frac{1}{z_0}$ (this corresponds to
$\alpha +\frac{\pi}{2}$ being also a solution if so is $\alpha$, and
this is related to the fact that (\ref{cond1}) and (\ref{cond2}) are 
symmetric under the interchange of 0 and 1). Since $p_i (1) = 
- p_i (-1)$, $p_i$ has a real root in $[-1,1]$. To simplify the
problem, we can extract a factor $z^2 +1$ from $p_i$, so we reduce
the problem to two polynomials of degree six, that have the same 
symmetry properties discussed above. If we introduce the variable
$w = (\frac{1}{z} -z)$ we can further reduce the two polynomials of
degree six to two polynomials $g_i (w) , i=1,2$ of degree three, given
by 
\begin{equation}
\label{gw}
g_i (w) = A_i  w^3 + B_i w^2 + (C_i + 2 A_i) w + (D_i + B_i),
\end{equation}
with the property that if $w$ is a root of $g_i$, the corresponding $z$'s
given by $w = (\frac{1}{z} -z)$ (which are real if and only if $w$ is real) 
are roots 
of $p_i$. So we reduced the problem to finding a common real root of $g_1$ 
and $g_2$. The resultant \cite{result} of the two polynomials $g_1$ and 
$g_2$ is a function of $\zeta$ and takes the form
\begin{equation}
\label{result}
{\mathrm{Res}}(g_1 , g_2) (\zeta)  = \sum_{k=0}^{4} (r_k \cos [(2 + 4 k) \zeta]
 + s_k \sin [(2 + 4 k) \zeta]),
\end{equation}
where $r_k$ and $s_k$ are polynomials on the coefficients of $\pstate$.
We can see that this resultant vanishes several times in $[0,2 \pi]$,
which is the range of $\zeta$, and this is useful because the resultant 
of two polynomials vanishes if and only if they have a common factor. This
falls short of saying that we can find $\zeta$ such that $g_1$ and $g_2$
have a common real root, because there is in principle the possibility
that the common factor is a polynomial of degree 2 irreducible over
the real numbers, so $g_1$ and $g_2$ have a common root but it is 
complex. However, after checking this procedure with many randomly
generated states, we found that the common factor \textit{always
corresponds to a real root}.

Let's assume that in fact, we can always find a value  $\zeta_0$ 
such that $p_1$ and $p_2$ have a common real root $z_0$. Then we
know that if we apply $U(\alpha_0,\zeta_0) \otimes \mathbf{1}
 \otimes \mathbf{1}$ (where $\alpha_0 
= \arctan (z_0)$) to $\pstate$, we obtain a state $| \psi' \rangle$
that satisfies (\ref{cond1}) and (\ref{cond2}). Then, we can choose a POVM that
satisfies (\ref{condpovm}), and we can be sure that both outcomes
of this POVM, when applied to  $| \psi' \rangle$, will
have the same values of $I_i , i=1,\ldots,5$. However, as we pointed
out before, this is still not enough to say that both outcomes are in the
same orbit. There's still the possibility that they are in orbits
that are conjugate to each other, since we are dealing with 
complex states, which are not LUeq to their conjugates. To decide
which one is the case, we can calculate the sign of the imaginary part of
$I_6$ for both outcomes. Unfortunately, the expression of $I_6$ for
both outcomes is too complicated and it's not possible
to extract the sign of the imaginary part analytically for
an arbitrary state, although it is very easy to compute it
numerically for a given state. We analysed randomly generated states, 
and found that we can 
\textit{always} find a value of $\zeta$ for which both outcomes are 
indeed in the same orbit (although there are other values of $\zeta$
for which the outcomes are in conjugate orbits). We will refer to states 
with this property as \textit{gate states}, since we can use them as a gate to leave
one orbit and move to another with probability 1.  

\section{The transformation in the space of orbits}

We can now use the results of the previous section to give a
characterization of the states that can be obtained from $\pstate$
by applying a 2-outcome deterministic POVM. Let us assume that 
the state $\pstate$ is a gate state. We will also assume that $a<b$ 
(if it's not, we apply
a bit flip on Alice's qubit, which interchanges the matrices
$T_0$ and $T_1$, and hence $a$ and $b$). We can use the invariants
evaluated for $\phizero$ (given by (\ref{invxy}))
to characterize the orbit of the outcomes. These equations are 
homogeneous of degree zero in $x$ and $y$, so we
can write them in terms of only one parameter $\lambda=\frac{y}{x}$
\begin{eqnarray}
\label{invz}
I_i (\lambda) & = & \alpha_i + \beta_i \frac{\lambda}{(a + b \lambda)^2} 
\ \ \ \ i=1,\ldots,4 
\nonumber \\
I_5 (\lambda) & = & \alpha_5 + \frac{\lambda(\beta_5 + \gamma_5 \lambda)}{(a + b 
\lambda)^3}, 
\end{eqnarray}
where 
\begin{eqnarray}
\label{coefinvz}
\alpha_1 & = &1 \ , \ \alpha_2=\alpha_3=\frac{Tr[(T_0 T_0^\dag)^2]}{a^2} 
\ , \ \alpha_4=0  \nonumber \\
\alpha_5 & = & \frac{Tr[(T_0 T_0^\dag)^3]}{a^3} \nonumber \\
\beta_1 & = & 2(Tr[T_0 T_1^\dag]Tr[T_1 T_0^\dag] - a b ) \nonumber \\
\beta_2 & = & 2(Tr[T_0 T_0^\dag T_1 T_1^\dag] - b \frac{Tr[(T_0 T_0^\dag)^2]}
{a}) \nonumber \\
\beta_3 & = & 2(Tr[T_0 T_1^\dag T_1 T_0^\dag] - b \frac{Tr[(T_0 T_0^\dag)^2]}
{a}) \nonumber \\
\beta_4 & = & I_4 (\pstate) \nonumber \\
\beta_5 & = & 3(Tr[T_0 T_1^\dag T_0 T_0^\dag T_1 T_0^\dag] - \frac{b
Tr[(T_0 T_0^\dag)^3]}{a}) \nonumber \\
\gamma_5 & = &  3(Tr[T_1 T_0^\dag T_1 T_1^\dag T_0 T_1^\dag] - \frac{b^2
Tr[(T_0 T_0^\dag)^3]}{a^2}). 
\end{eqnarray}
The range of $\lambda$ is $[1,+\infty)$ (when $a<b$), with $\lambda=1$ 
corresponding to no
transformation ($E_0 \propto \mathbf{1}$), so we have $I_i (\lambda=1) = 
I_i (\pstate)$, and $\lambda=+\infty$ corresponds to a projective
measurement ($y=1, x=0$), that destroys any 3-particle entanglement. From
(\ref{invz}) we can see that the set of orbits we can reach from $\pstate$
can be described as a one parameter family $\{I_i (\lambda)\}$ 
that corresponds to a curve
in the space of orbits, that starts at state $\pstate$ and ends on a state
that has no tripartite entanglement.

It is possible for some orbits to have more than one gate state. 
The values of the coefficients (\ref{coefinvz}) will be in general
different for different gate states. Since these coefficients determine
the curve $\{ I_i (\lambda) \}$, we will be able to transform to
different sets of orbits depending on which gate state we use.  
We can also reach a different
family of orbits if we let Bob or Charlie apply a deterministic
POVM instead of Alice. This is because the matrices $T_i$, are different 
for different parties, and so will give in general different gate states.  

If we fix the sign of the imaginary part of $I_6$, we can use
the invariants $\{ I_i, i=1,\ldots,5 \}$ as coordinates for
the orbits. All the previous results can be summarized in the
following picture. Every point in this space (which represents
the orbit of some state $\pstate$), is the starting
point  of a finite number of curves, each representing
a set of orbits that can be obtained from $\pstate$ with
probability 1 with a local 2-outcome POVM.

More orbits can be reached if several rounds of deterministic POVMs
are allowed. The general protocol will be something like this: 
(i) starting with the state $\pstate$, Alice
applies a local unitary to transform it into a gate state;
(ii) she applies a POVM
on her part of the system, that satisfies (\ref{condpovm});
(iii) according to the outcome she obtains, she communicates to Bob 
and Charlie the state $|\psi ' \rangle$ they are sharing after the
measurement;(iv) they decide which one will apply the next POVM and
repeat the same steps, now starting with the state  $|\psi ' \rangle$. 
A simplified pictorial representation of this transformation is given
in Figure 1. 
\begin{figure}
\setlength{\unitlength}{1mm}
\begin{picture}(85,70)

\thinlines
\put(5,5){\vector(0,1){55}}
\put(5,5){\vector(1,0){70}}
\put(0,55){$I_5$}
\put(70,2){$I_4$}
\put(66,53){\circle*{1.5}}
\put(49,38){\circle*{1.5}}
\put(26,23){\circle*{1.5}}
\put(69,51){$| \psi \rangle$}
\put(52,36){$| \psi' \rangle$}
\put(29,21){$| \psi'' \rangle$}
\thicklines
\qbezier(49,38)(54,48)(66,53)
\qbezier(26,23)(34,32)(49,38)

\qbezier[10](52,50)(58,53)(66,53)
\qbezier[10](60,43)(62,49)(66,53)

\qbezier[10](45,26)(46,30)(49,38)
\qbezier[10](38,27)(42,33)(49,38)

\qbezier[10](13,18)(15,20)(26,23)
\qbezier[10](17,12)(19,16)(26,23)

\end{picture}
\caption{Transformation of states in the space of orbits. Full lines
represent a particular transformation that takes $\pstate$ to $| \psi' 
\rangle$ and then to  $| \psi'' \rangle$. Dotted lines correspond to
other possible choices for deterministically transforming the states.}
\end{figure}
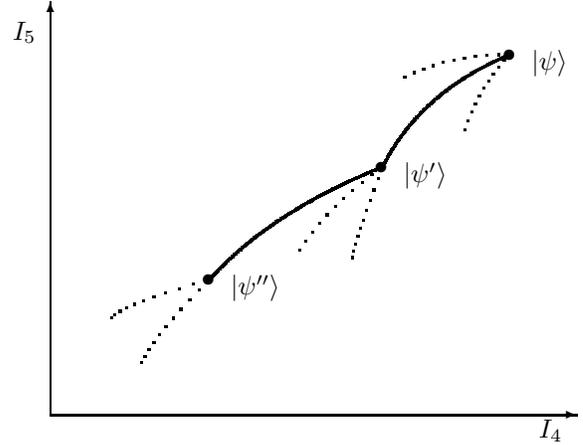
The transformation occurs in the 5-dimensional space
defined by the invariants $I_i$, but
for simplicity, we represent only two of them ($I_4$ and $I_5$). We start
with a gate state $\pstate$ and we apply a deterministic 2-outcome POVM (with
some parameter $\lambda_0$), 
that transforms it into state $| \psi' \rangle$. The line connecting
$\pstate$ and $| \psi' \rangle$ represents all the orbits that can be reached
from $\pstate$ by applying a POVM with parameter $\lambda$ between 1 and 
$\lambda_0$. The dotted lines originating at $\pstate$ represent the 
set of orbits that can be reached from the same orbit, but using a gate
state different from $\pstate$ (which is in the same orbit as $\pstate$
so it's represented by the same point in the plot). In the 
actual space of orbits, these curves extend until they reach an orbit
that represents a state with no 3-particle entanglement, that corresponds
to the point where the POVM becomes a projective measurement (i.e., 
$\lambda = +\infty$). For clarity, we are only plotting the beginning of 
these curves. After deterministically transforming $\pstate$ into
$| \psi' \rangle$, the parties can choose again from several gate states
to apply the next POVM. This will determine which party will apply this
POVM, because in general, a state is a gate state only for a particular
party. In the figure, the full line represents a POVM that transforms
$| \psi' \rangle$ into  $| \psi'' \rangle$, while again, the dotted lines
correspond to other possible deterministic transformations that can
be applied to  $| \psi' \rangle$. By applying many 
deterministic POVMs with different parameters, we can reach many different
orbits.

\section{Transformation of the state ${\mathbf |GHZ\rangle = \frac{1}{\sqrt{2}} 
(|000\rangle +|111\rangle)}$}

As an example of the use of 2-outcome deterministic POVMs, we will now study the 
particular case of the $|GHZ\rangle$ state. As it
was mentioned in Section II, this state belongs to the subclass of states
that satisfy ${\mathrm{Re}}(\Omega)=0$. We will show that it can be transformed
with probability 1, into any real state in that subclass.

First, we need to identify the real states that satisfy ${\mathrm{Re}}(\Omega)=0$.
Then clearly we must have that either $\Omega(\pstate)$ is zero or pure imaginary.
In the former case, this means that $\langle \mu | \nu \rangle = 0$, and
then decomposition (\ref{ghzdec}) takes one of the following forms:     
\begin{eqnarray}
\label{ghzorbit}
\mu |0 0 0 \rangle & + & \nu  |1 \rangle 
 |\varphi \rangle  |\varphi' \rangle \nonumber \\
\mu |0 0 0 \rangle & + & \nu  |\varphi \rangle 
 |1 \rangle  |\varphi' \rangle \nonumber \\
\mu |0 0 0 \rangle & + & \nu |\varphi \rangle 
 |\varphi' \rangle  |1 \rangle. 
\end{eqnarray}
If $\Omega(\pstate)$ is pure imaginary, then the only case in which
$\pstate$ is actually a real state is the case in which $\mu = \nu$, as 
discussed in Section II. In this case, the state takes the form
\begin{equation}
\label{compreal}
\frac{1}{\sqrt{2}} (|0 0 0 \rangle \pm  i    
 |\varphi \rangle  |\varphi' \rangle |\varphi'' \rangle),
\end{equation}
where none of the states in the
second term can be equal to $|0\rangle$ or $|1\rangle$ (otherwise it 
could be transformed into a real state by a local unitary), and we
obtain $\mu=\frac{1}{\sqrt{2}}$ by imposing normalization of the state.
The two states in (\ref{compreal}) (corresponding to the two
possible signs of the second term) are LUeq to each other.

Since the GHZ state is symmetric under a permutation of the parties,
it is clear that if we find a protocol that transforms it into the first state 
in (\ref{ghzorbit}), then we can also transform it into the 
other two. In this section we will use the results of Section III to 
explicitly construct protocols that transform the GHZ state into the state 
\begin{equation}
\label{real}
|\phi\rangle = \mu |0 0 0 \rangle + \nu  |1 \rangle 
 |\varphi \rangle  |\varphi' \rangle,
\end{equation}
or the state
\begin{equation}
\label{comp}
\frac{1}{\sqrt{2}} (|0 0 0 \rangle +  i    
 |\varphi'' \rangle  |\varphi \rangle |\varphi' \rangle),
\end{equation}
for all allowed values of $\mu, \nu, |\varphi\rangle,|\varphi'\rangle$ 
and $|\varphi''\rangle$. 
These protocols will be divided into three steps. First, Charlie applies
a local deterministic POVM that transforms $|GHZ\rangle$ into $ \frac{1}{\sqrt{2}} 
(|000\rangle +|11\rangle | \varphi'\rangle)$. Then, Bob applies another local
POVM that takes the state to $ \frac{1}{\sqrt{2}} (|000\rangle +|1\rangle
|\varphi\rangle |\varphi'\rangle)$. Finally, Alice applies the last POVM,
which she can choose to take the state to $\mu |000\rangle +\nu|1\rangle
|\varphi\rangle |\varphi'\rangle$ or $\frac{1}{\sqrt{2}} (|0 0 0 \rangle +  i    
 |\varphi'' \rangle  |\varphi \rangle |\varphi' \rangle)$.

{\textit{Step 1.}} The $T_i$ matrices for the GHZ state are given by
\begin{equation}
\label{Tghz}
 T_0 = \left( \begin{array}{cc} \frac{1}{\sqrt{2}} & 0 \\ 0 & 0 \end{array}
        \right) \ \  , \ \ 
 T_1 = \left( \begin{array}{cc} 0 & 0 \\ 0 & \frac{1}{\sqrt{2}} \end{array}
        \right),
\end{equation}
and they have the same form for all parties. If Charlie applies a local
unitary $U$ on its qubit, where
\begin{equation}
\label{U}
U = \frac{\sqrt{2}}{2}  \left( \begin{array}{cc} 1& 1 \\ -1 & 1 \end{array}
        \right),  
\end{equation}
the $T'_i$ matrices for the state $|\psi'\rangle={\mathbf 1} \otimes 
{\mathbf 1} \otimes U |GHZ\rangle$ are
\begin{equation}
\label{UT}
 T'_0 = \frac{1}{2}\left( \begin{array}{cc} 1 & 0 \\ 0 & 1 \end{array}
        \right) \ \ , \ \
 T'_1 = \frac{1}{2}\left( \begin{array}{cc} -1 & 0 \\ 0 & 1 \end{array}
        \right).
\end{equation}
It is very easy to see that these matrices satisfy equation (\ref{cond1}),
so the state $|\psi'\rangle$ is a gate state. Thus, Charlie can apply a 
deterministic POVM to it. In particular, this state satisfies $b'= a' 
= Tr[T_0^\dag T_0]=\frac{1}{2}$, so according to equation (\ref{condpovm}) we
have $y=1-x$, so Charlie can apply a deterministic POVM of the form
\begin{equation}
\label{POVMC}
 E_0 = \left( \begin{array}{cc} \sqrt{x} & 0 \\ 0 & \sqrt{1-x} \end{array}
        \right) \ \  , \ \ 
  E_1 = \left( \begin{array}{cc} \sqrt{1-x} & 0 \\ 0 & \sqrt{x} \end{array}
        \right),
\end{equation}
where $x \in [\frac{1}{2},1)$. The normalized state corresponding to the 
outcome zero is
\begin{eqnarray}
\label{outzero}
|\phi_0\rangle & = & \sqrt{2}({\mathbf 1} \otimes 
{\mathbf 1} \otimes E_0) |\psi'\rangle   \nonumber \\
               & = &  \sqrt{2}({\mathbf 1} \otimes 
{\mathbf 1} \otimes E_0 U ) |GHZ\rangle   \nonumber \\
               & = & |00\rangle \,(E_0 U |0\rangle) + |11\rangle \,(E_0 U |1\rangle)
\nonumber \\
               & = & \langle 0|U^\dag E_0^\dag E_0 U |0\rangle^{\frac{1}{2}}\,
 |00\rangle |0'\rangle  + \nonumber \\
& & \ \ \ \ \ \ \ \ \ \ + \langle 1|U^\dag E_0^\dag E_0 U |1\rangle^{\frac{1}{2}}\,
 |11\rangle |1'\rangle, 
\end{eqnarray}
where 
\begin{equation}
\label{01prime}
|0'\rangle = \frac{E_0 U |0\rangle}{ \langle 0|U^\dag E_0^\dag E_0 U 
|0\rangle^{\frac{1}{2}}} \ \ , \ \ 
|1'\rangle = \frac{E_0 U |1\rangle}{ \langle 1|U^\dag E_0^\dag E_0 U 
|1\rangle^{\frac{1}{2}}},
\end{equation}
are normalized states. A straightforward calculation shows that 
$\langle 0|U^\dag E_0^\dag E_0 U |0\rangle = \langle 1|U^\dag E_0^\dag E_0 U 
|1\rangle = \frac{1}{2}$, so we can write
\begin{equation}
|\phi_0\rangle = \frac{1}{\sqrt{2}}(|00\rangle |0'\rangle + |11\rangle |1'\rangle).
\end{equation}
This state can be taken to the canonical form (\ref{ghzdec}) by letting
Charlie apply a local (real) unitary on his qubit, that takes the state
$|0'\rangle$ into $|0\rangle$, and $|1'\rangle$ into $|\varphi'\rangle
= \cos \delta' |0\rangle + \sin \delta' |1\rangle$. Thus, $\langle 0|
\varphi'\rangle=\langle 0'|1'\rangle$ and then we have
\begin{eqnarray}
\cos \delta' & = & \langle 0' | 1' \rangle \nonumber \\
             & = & 2 \langle 0 | U^\dag E_0^\dag E_0 U |1\rangle \nonumber \\
             & = & 2 x-1.
\end{eqnarray}
We can see that for any $\delta' \in (0,\frac{\pi}{2}]$, we can find
$x \in [\frac{1}{2},1)$ that satisfies this equation. This means that
we can transform $|GHZ\rangle$ into $\frac{1}{\sqrt{2}}(|000\rangle+ |11\rangle 
|\varphi'\rangle)$ with probability 1, for any $|\varphi'\rangle$.

{\textit{Step 2.}} In this step Bob applies a deterministic
POVM to transform the state \hbox{$|\phi\rangle = \frac{1}{\sqrt{2}}(|000\rangle+ 
|11\rangle |\varphi'\rangle)$} into $\frac{1}{\sqrt{2}}(|000\rangle+ |1\rangle 
|\varphi\rangle| \varphi'\rangle)$. The $T_i$ matrices for $|\phi\rangle$
from Bob's point of view, are given by
\begin{equation}
 T_0 = \left( \begin{array}{cc} \frac{1}{\sqrt{2}} & 0 \\ 0 & 0 \end{array}
        \right) \ \  , \ \ 
 T_1 = \left( \begin{array}{cc} 0 & 0 \\  \frac{1}{\sqrt{2}} \cos \delta' & 
\frac{1}{\sqrt{2}} \sin \delta' \end{array}
        \right).
\end{equation}
First, Bob applies the local unitary $U$ from (\ref{U}) to his qubit, 
obtaining the state $|\phi'\rangle = {\mathbf 1} \otimes U \otimes {\mathbf 1} 
|\phi\rangle$, characterized by matrices $T'_i$ given by
\begin{equation}
 T'_0 = \frac{1}{2}\left( \begin{array}{cc} 1 & 0 \\ \cos \delta' & \sin \delta' 
\end{array}
        \right) \ \  , \ \ 
 T'_1 = \frac{1}{2}\left( \begin{array}{cc} -1 & 0 \\  \cos \delta' & 
\sin \delta' \end{array}
        \right).    
\end{equation}
Again, it is easy to show that $T'_i$ satisfy (\ref{cond1}), so $|\phi'\rangle$
is a gate state. We also have that $a' = Tr[T'_0 {T'_0}^\dag] = \frac{1}{2}$, so 
Bob can apply the POVM of equation (\ref{POVMC}) to his qubit and obtain two
outcomes in the same orbit. We can apply the same analysis we did in Step 1
to the outcome $|\chi_0\rangle$ of Bob's POVM, and show that
\begin{equation}
|\chi_0\rangle = \frac{1}{\sqrt{2}}(|0\rangle |0'\rangle |0\rangle + |1\rangle
|1'\rangle |\varphi'\rangle)
\end{equation}
where the normalized states $|0'\rangle$ and $|1'\rangle$ are also given by
(\ref{01prime}). It should be clear from Step 1 that, again, we can choose $x$ 
and a suitable local
unitary on Bob's qubit to transform this state into
\begin{equation}
|\chi\rangle =   \frac{1}{\sqrt{2}}(|000\rangle + |1\rangle
|\varphi\rangle |\varphi'\rangle)
\end{equation}
for any $|\varphi\rangle = \cos \delta |0\rangle + \sin \delta |1\rangle$,
with $\delta \in (0,\frac{\pi}{2}]$.

{\textit{Step 3.}} Now Alice has to choose between two local POVMs
depending on whether she wants to obtain (\ref{real}) or (\ref{comp}).
Consider first the case in which she wants to transform $|\chi\rangle$ into 
$\mu |000\rangle + \nu |1 \varphi \varphi'
\rangle$. The $T_i$ matrices for $|\chi\rangle$ from Alice's point of view are
\begin{equation}
 T_0 =  \frac{1}{\sqrt{2}}\left( \begin{array}{cc} 1 & 0 \\ 0 & 0 \end{array}
        \right) \  , \ 
 T_1 =  \frac{1}{\sqrt{2}}\left( \begin{array}{cc} \cos \delta  \cos \delta' & 
\cos \delta \sin \delta' \\  \sin \delta \cos \delta' & 
\sin \delta \sin \delta' \end{array}
        \right).
\end{equation}
These matrices already satisfy equation (\ref{cond1}) and since 
\hbox{$a=Tr[T_0 T_0^\dag] = \frac{1}{2}$}, Alice can apply the deterministic
POVM given by (\ref{POVMC}). The state corresponding to outcome zero is
\begin{eqnarray}
\label{prot1}
|\xi\rangle & = & \sqrt{x} |000\rangle + \sqrt{1-x} |1 \varphi \varphi'\rangle
\nonumber \\
            & = & \mu  |000\rangle + \nu |1 \varphi \varphi'\rangle,
\end{eqnarray}
where we set $\mu=\sqrt{x}$ and $\nu=\sqrt{1-x}$.
Since $x \in [\frac{1}{2},1)$, we have $\mu \geq \nu$. The state in (\ref{prot1})
is the same as in (\ref{real}). 

Consider now the case in which Alice wants to obtain $\frac{1}{\sqrt{2}} 
(|0 0 0 \rangle +  i |\varphi'' \rangle  |\varphi \rangle |\varphi' \rangle)$
from $|\chi\rangle$. In this case we can construct the appropriate POVM
$\{ A_0 , A_1 \}$ by inspection. If $|\varphi'' \rangle = \cos \delta'' |0
\rangle + \sin \delta'' |1\rangle$, we define
\begin{equation}
 A_0 =  \frac{1}{\sqrt{2}}\left( \begin{array}{cc} 1 & i \cos \delta'' \\ 0 & 
i \sin \delta'' \end{array}
        \right) \  , \ 
 A_1 =  \frac{1}{\sqrt{2}}\left( \begin{array}{cc} 1 & - i \cos \delta'' \\ 0 & 
- i \sin \delta'' \end{array}
        \right).
\end{equation}
It is easy to verify that they satisfy $A_0^\dag A_0 + A_1^\dag A_1 = {\mathbf{1}}$,
and that the probabilities of both outcomes are equal to $\frac{1}{2}$. The normalized
state that corresponds to outcome zero is
\begin{equation}
\label{out0}
\frac{1}{\sqrt{2}} 
(|0 0 0 \rangle +  i |\varphi'' \rangle  |\varphi \rangle |\varphi' \rangle),
\end{equation}
while the one corresponding to outcome 1 is just the complex conjugate 
of (\ref{out0}). But we know that these two states are actually in the
same orbit, so we can transform outcome 1 into (\ref{out0}) by local
unitaries, so we obtain (\ref{out0}) with probability 1. The state in (\ref{out0}) 
is the same as in (\ref{comp}). This concludes the protocol.

Note that all three steps
involve only local unitaries and deterministic POVMs, so these protocols allow
Alice, Bob and Charlie to transform the GHZ state into any other real state 
that belongs to the subclass defined by ${\mathrm{Re}}(\Omega)=0$
with probability 1, using only local operations and classical communication.
This is then a complete characterization of the real states that can
be obtained from the GHZ state, since by Theorem 1 we know that we
cannot reach real states that belong to a different subclass. It is 
interesting to note that it does
not seem to be that easy to find a deterministic protocol to transform the GHZ state into
any complex state in the same subclass. Whether this is actually possible
is still an open question.

\section{Summary and Conclusions}

In this paper, we studied the properties of deterministic LOCC
transformations of 3-qubit pure states with tripartite entanglement.
We showed that the set of states in the GHZ class breaks into an infinite 
number of disjoint subclasses, characterized by the real part of a complex function
$\Omega(\pstate)$.
Two states that belong to different subclasses cannot be transformed one into
the other with probability one, by means of local operations and 
classical communication.  This quantity is not only invariant under 
deterministic transformations, but it also must be conserved by any local
POVM that is part of a deterministic protocol. This imposes a strong
constraint on the POVMs that can be used for deterministically transforming
a given state. 

It is interesting to point out that the invariance of ${\mathrm{Re}}(\Omega)$ under
deterministic LOCC transformations (and its invariance under any local POVM
that is part of such a transformation), follows from the invariance of $\Omega$ 
under local unitaries and the very particular form of
equation (\ref{phaseave}). In the language of entanglement monotones, we can
say that ${\mathrm{Re}}(\Omega)$ is both an increasing and decreasing
entanglement monotone. Any function
of the states that is invariant under local unitaries and satisfies an equation 
like (\ref{phaseave}) for an arbitrary local POVM, 
will be invariant under deterministic
LOCC protocols, and hence will break the set of states into inequivalent classes
that will be labeled by that function. This will be true even in the multipartite 
case, so identifying quantities with these properties could be very useful in
the study of deterministic transformations of entanglement.

We also discussed the case of a deterministic 2-outcome POVM. 
We showed that for this POVM to exist, both the state and
the parameters of the POVM have to satisfy certain polynomial
conditions. In particular the coefficients of the state have
to satisfy two polynomial constraints. To be able to apply
a deterministic POVM to a given state, we need to find a local
unitary that will transform our original state into another
state that satisfies the two constraints. For
real states, the problem actually simplifies and only one
constraint has to be satisfied. In this case, it was proven
in general that the necessary local unitary could be found, allowing
us to apply a local 2-outcome POVM that would send the state to
some other orbit with probability 1.
For complex states we found
some analytical evidence that the unitary could be found, but a 
rigorous proof of this
fact is still an open problem. However, it is important to stress
that of all random numerical examples analysed, the algorithm discussed
in Section II never failed to find a gate state for complex states.
We also discussed how several rounds of POVMs and local unitaries applied 
in sequence
by all the parties allow us to reach a bigger set of orbits than the
one we get from only one POVM. There is a lot of freedom in choosing
the order in which the parties apply a POVM and which POVM they
choose. Although it is in general difficult to study this
procedure analytically, in order to characterize the set of states that can be
obtained from $\pstate$ (except for states with high symmetry like
the GHZ state), 
a numerical analysis is easy to implement, and can be used to study
general properties of this set, that could help us to have a better 
understanding of deterministic transformations. 

Finally, we combined the two main results of this paper to give a complete
characterization of the real states that can be obtained from the GHZ state
with probability 1. First we used the results of Section II to characterize
the subclass of states that could in principle be obtained deterministically 
from it, and then we constructed  an 
explicit protocol that 
allows the three parties to transform the GHZ state into any 
real state in that subclass. Finding a protocol to transform it to a complex
state in the same subclass does not seem to be as easy, and thus whether this
transformation is possible or not is still an open question.

\section{Acknowledgements}

I would like to thank my advisor John Preskill for his support during
this research and for many useful suggestions. I am also very grateful
to Bob Gingrich and Guifr\'e Vidal for many useful discussions, 
to Pablo Parrilo for his very useful technical advice on polynomials,
and to John Cortese for his suggestions and comments to improve the
manuscript. This work has been supported in part by the National Science
Foundation under Grant No. EIA-0086038.

\appendix
\section{Solution of $\mathbf{I}_{\lowercase{\mathit{i}}} (\lowercase{x},
\lowercase{y}) 
= \mathbf{I}_{\lowercase{\mathit{i}}} 
(1-\lowercase{x},1-\lowercase{y})$}

We want to know under which conditions does (\ref{equalinv}) have 
a nontrivial solution (i.e., $x \neq y$ and $x,y \neq 0,1$). 
We will consider only states that have 3-particle entanglement, which
means that $ a,b \neq 0,1$. First, let us note that we can write
$I_1 (x,y)$ as
\begin{equation}
\label{I1}
  I_1 (x,y) = 1 + \frac{2 x y (Tr[T_0 T_1^\dag]Tr[T_1 T_0^\dag] - a b)}{
                       (a x + b y)^2},
\end{equation}
where $(Tr[T_0 T_1^\dag]Tr[T_1 T_0^\dag] - a b) \neq 0$ if $\pstate$
has 3-particle entanglement. Then $I_1 (x,y) = I_1 (1-x,1-y)$ has a 
solution if and only if
\begin{equation}
\label{xycond1}
\frac{x y}{(a x + b y)^2} = \frac{(1-x)(1-y)}{(a (1-x) + b (1-y))^2},
\end{equation}
which is the same as
\begin{equation}
\label{xycond2}
a^2 x (1-x) = b^2 y (1-y).
\end{equation}
This also implies that $I_4 (x,y) = I_4 (1-x,1-y)$. Both $I_2$
and $I_3$ have the form
\begin{equation}
\label{I23}
I_i (x,y) = \frac{F_0 x^2 + F_1 y^2 + 2 C_i x y}{(a x + b y)^2}  \ \ \ i=2,3.
\end{equation}
We can use (\ref{xycond1}) to write $I_i (x,y) = I_i (1-x,1-y), i=2,3$ as
\begin{equation}
\label{I232}
 \frac{F_0 + F_1 z^2}{(a + b z)^2} =\frac{F_0 + F_1 w^2}{(a + b w)^2},
\end{equation}
where we introduced the variables $z = \frac{y}{x}$ and 
$w = \frac{(1-y)}{(1-x)}$. From (\ref{xycond2}) we see that these 
variables are not independent, and satisfy the condition 
$z w = (\frac{a}{b})^2$. Furthermore, both $z$ and $w$ are positive,
since $x$ and $y$ are between 0 and 1. If we expand (\ref{I232}) and
use the relationship between $z$ and $w$, we have
\begin{equation}
(F_0 b^2 - F_1 a^2)(z \frac{(a^2 + b^2)}{a^2} + 2 \frac{a}{b}) = 0,
\end{equation}
and since $z$ has to be positive (and $a$ and $b$ are positive), we
have the condition
\begin{equation}
\label{con1}
a^2 F_0 = b^2 F_1,
\end{equation}
which is equation (\ref{cond1}).

To study the equation $I_5 (x,y) = I_5 (1-x,1-y)$ we can assume
that both (\ref{con1}) and (\ref{xycond2}) are satisfied, since
we are looking for a \textit{simultaneous} solution of 
(\ref{equalinv}). Let $\mu = I_5 (x,y)$. Introducing $z=\frac{y}{x}$
and using (\ref{invxy}) we can write
\begin{equation}
G_{00} + 3 G_{01} z + 3 G_{10} z^2 + G_{11} z^3 = \mu (a + b z)^3,
\end{equation}
where $G_{ij} = Tr[ T_i T_j^\dag T_i T_i^\dag T_j T_i^\dag ]$, and we
can expand this into
\begin{eqnarray}
\label{cubic}
(G_{00} - \mu a^3) & + & 3 (G_{01} - \mu a^2 b) +3(G_{10} - \mu a b^2) z^2 
+ \\
 & & \ \ \ \ \ +(G_{11} - \mu b^3) = 0.
\end{eqnarray}
A root of this cubic polynomial represents an operator of a POVM for
which the value of $I_5$ for the outcome of that operator is $\mu$. 
We are looking for two operators whose outcomes have the same
value of $I_5$, but that also satisfy equation (\ref{xycond2}). That
is the same as finding two roots $z_0$ and $z_1$ of (\ref{cubic}), that
satisfy the condition 
\begin{equation}
\label{rootcond}
z_0 z_1 = \frac{a^2}{b^2}.
\end{equation}
Let $z_2$ be the third root of (\ref{cubic}). From elementary algebra 
we know that the product of the three roots is equal to minus the 
quotient of the independent and the cubic coefficients, so we can write   
\begin{equation}
\label{3rootcond}
z_0 z_1 z_2 = - \frac{G_{00} -\mu a^3}{G_{11} -\mu b^3} =
- \frac{a^3}{b^3} \frac{\frac{G_{00}}{a^3} - \mu}{\frac{G_{11}}{b^3} - \mu} .
\end{equation}
Using (\ref{con1}) and the Cayley-Hamilton theorem, it can be 
shown that
\begin{equation}
\frac{G_{00}}{a^3} = \frac{G_{11}}{b^3},
\end{equation}
so (\ref{3rootcond}) reduces to
\begin{equation}
z_0 z_1 z_2 = - \frac{a^3}{b^3}.
\end{equation}
If we want (\ref{rootcond}) to be satisfied we need $z_2 =-\frac{a}{b}$.
If we plug this into (\ref{cubic}), we find that $z_2$ is actually
a root if and only if
\begin{equation}
b\ G_{01} = a F_{10},
\end{equation}
which is equation (\ref{cond2}).
There is one more detail we need to check. We need $z_0 =
\frac{y}{x}$ and $z_1=\frac{1-y}{1-x}$
to be positive numbers, because $x$ and $y$ are between 0 and 1,
and only one of them should be greater than 1 (which can be seen from
their explicit form in terms of $x$ and $y$). We know that the other
root $z_2=-\frac{a}{b}$ is negative, so the condition for only
one of them to be greater than 1 can be written
\begin{equation}
(z_0 -1)(z_1 -1)(z_2 -1) > 0.
\end{equation}
Expanding this inequality we get
\begin{equation}
z_0 z_1 z_2 -(z_0 z_1 + z_0 z_2 + z_1 z_2) + (z_0 + z_1 + z_2) -1 >0.
\end{equation}
All the symmetric polynomials on the roots of a polynomial equation
can be written in terms of the coefficients of that polynomial, so
we can rewrite this inequality as
\begin{equation}
-\frac{(G_{00} - \mu a^3)}{(G_{11} - \mu b^3)} - 
3\frac{(G_{01} - \mu a^2 b)}{(G_{11} - \mu b^3)}-
3 \frac{(G_{10} - \mu a b^2)}{(G_{11} - \mu b^3)} -1 > 0. 
\end{equation}
Expanding this and using $a+b=1$ we get
\begin{equation}
G_{00}+3\ G_{01} +3\ G_{10} + G_{11} > \mu.
\end{equation}
But the left hand side is just the value of $I_5$ for the
state $\pstate$, while $\mu$ is the value of $I_5$ for the
transformed state $\phizero$ (or $\phione$). So this 
condition is telling us that under a deterministic 
2-outcome POVM, $I_5$ behaves monotonically, even though
it is not an entanglement monotone in general.

\bibliographystyle{prsty}
\bibliography{biblio}

\begin{thebibliography}{10}
\bibitem{svd}
This decomposition can be understood in the following way: $A_i$ are 
positive semidefinite operators, and by performing a singular value decomposition
we know that there are unitary matrices $V_i$ and $U_i$ such that
$V_i^\dag A_i U_i^\dag$ are positive semidefinite diagonal matrices.
The matrices $U_i$ can be used to diagonalize the hermitian matrices 
$A_i^\dag A_i$. But since we have the constraint that $A_0^\dag A_0 +
A_1^\dag A_1 = {\mathbf{1}}$, and $ {\mathbf{1}}$ is already a diagonal matrix,
it is easy to see that we can take $U_0 = U_1 = U$.
\end{thebibliography}

\end{document}